\documentclass[
    ,final            
  ]
  {aipproc}

\layoutstyle{8x11double}

\begin{document}

\newcommand{\chandra}{{\it Chandra}}
\newcommand{\spitzer}{{\it Spitzer}}

\def\ga{\ifmmode\stackrel{>}{_{\sim}}\else$\stackrel{>}{_{\sim}}$\fi}
\def\la{\ifmmode\stackrel{<}{_{\sim}}\else$\stackrel{<}{_{\sim}}$\fi}
\def\farcm{\hbox{$.\mkern-4mu^\prime$}}
\def\farcs{\hbox{$.\!\!^{\prime\prime}$}}

\title{High Energy Studies of Pulsar Wind Nebulae}

\classification{01.30.Cc;95.85.Hp;95.85.Bh;
95.85.Nv;95.85.Pw;97.60.Gb;98.38.Mz}

\keywords      {Pulsar Wind Nebulae; Pulsars; Supernova Remnants}

\author{Patrick Slane}{
  address={Harvard-Smithsonian Center for Astrophysics}
}

%

\begin{abstract}
The extended nebulae formed as pulsar winds expand into their
surroundings provide information about the composition of the winds,
the injection history from the host pulsar, and the material into which
the nebulae are expanding. Observations from across the electromagnetic
spectrum provide constraints on the evolution of the nebulae, the density
and composition of the surrounding ejecta, the geometry of the systems,
the formation of jets, and the maximum energy of the particles in the
nebulae. Here I provide a broad overview of the structure of pulsar wind
nebulae, with specific examples that demonstrate our ability to constrain
the above parameters. The association of pulsar wind nebulae with extended
sources of very high energy gamma-ray emission are investigated,
along with constraints on the nature of such high energy emission.
\end{abstract}

\maketitle


It has long been known that the Crab Nebula is produced by the wind
from a young, energetic pulsar whose spin-down power manifests
itself as a synchrotron-emitting bubble of energetic particles.
Optical observations reveal a network of filaments where the nebula
has swept up surrounding ejecta from the progenitor star, and an
inner shock region where the pulsar wind joins the slower flow in
the nebula. Observations from the radio band to the very high energy
(VHE) $\gamma$-ray band reveal the presence of energetic particles
swimming through a bubble of magnetic field and photons, some
self-generated, resulting in copious synchrotron and inverse-Compton
emission. (For a recent review on the Crab nebula, see Hester 2008.)
More recently, it has become clear that this basic structure
holds for all pulsar wind nebula (PWNe), and observed variations
in their properties provide us with information about differences
in their conditions and evolutionary states.

The diagram shown in Figure~1 (inset) illustrates the main points of the
most basic picture for a PWN: in an inner zone the wind flows away
from the neutron star (NS) with Lorentz factor $\gamma \sim 10^6$;
at a distance $R_w$ from the NS the wind passes through a termination
shock, decelerating the flow while boosting particle energies by
another factor of $\ga 10^3$; and beyond $R_w$ energetic electrons
in the wind radiate synchrotron emission in the wound-up toroidal
magnetic field to form the PWN which is confined at a radius $R_{PWN}$
by the inertia of the SN ejecta or the pressure of the interior of
a surrounding SNR. A detailed theoretical framework, incorporating
particle injection and diffusion, magnetic field evolution and
radiative and adiabatic losses, has been constructed within this
picture, allowing us to successfully predict and explain some basic
PWN properties (Reynolds \& Chevalier 1984; Kennel \& Coroniti
1984).  For a more detailed review on the structure and evolution
of PWNe, see Gaensler \& Slane (2006).

\begin{figure}[t]
  \includegraphics[height=0.3\textheight]{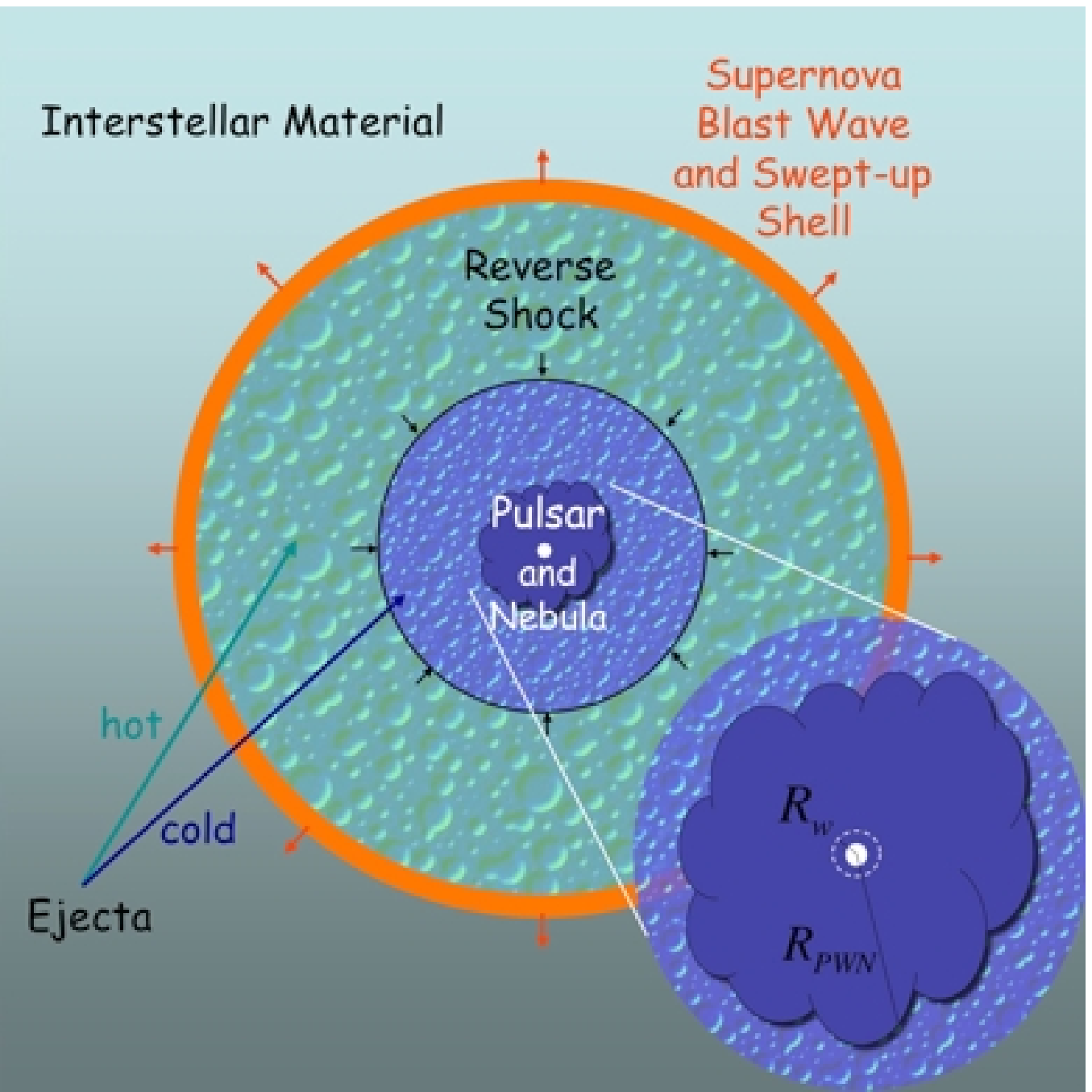}
  \includegraphics[height=0.3\textheight]{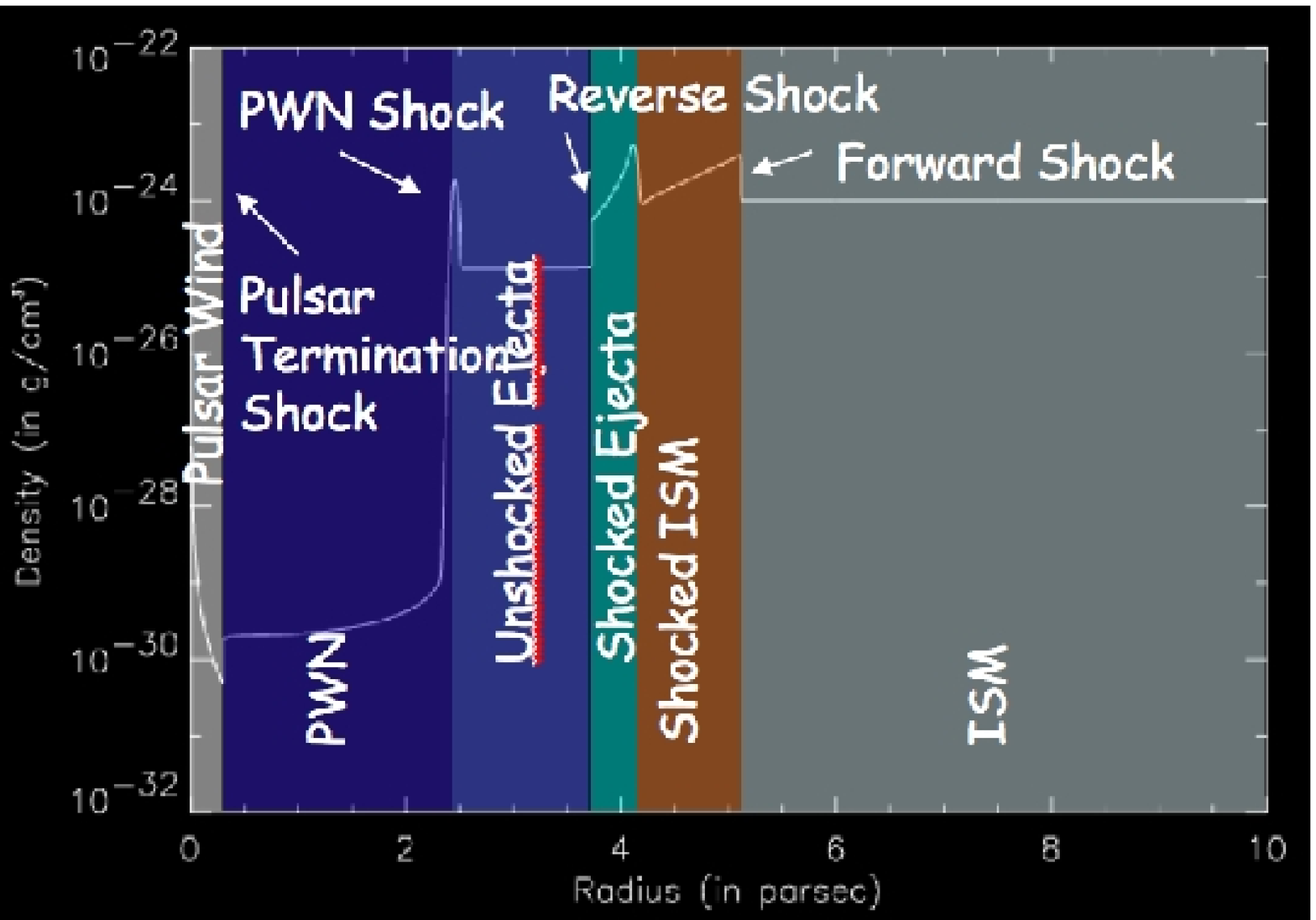}
  \caption{
Left: Conceptual diagram of a PWN within an SNR, illustrating the
central pulsar and nebula, surrounding cold ejecta, shock-heated ejecta,
and surrounding medium. 
Right: Density structure within a composite SNR, identifying
the discrete zones illustrated in the schematic diagram. The position
of the forward and reverse shocks that heat the CSM/ISM and ejecta
are shown, along with that of the PWN shock that expands into cold 
ejecta, and that of the pulsar wind termination shock. See text for
a detailed description.
}
\end{figure}

\section{Broadband Emission from PWNe}

The broadband emission from a PWN is determined by the power and
spectrum of particles injected by the pulsar, as well as the
environment into which the PWN expands. The input luminosity is
e.g. Pacini \& Salvati 1973)
\begin{equation}
\dot{E} = \dot{E}_0 \left( 1 + \frac{t}{\tau_0}
\right)^{-\frac{(n+1)}{(n-1)}},
\end{equation}
where
\begin{equation}
\tau_0 \equiv \frac{P_0}{(n-1)\dot{P}_0}
\end{equation}
is the initial spin-down time scale of the pulsar.  Here $\dot{E_0}$
is the initial spin-down power, $P_0$ and $\dot{P}_0$ are the initial spin
period and its time derivative, and $n$ is the so-called "braking
index" of the pulsar ($n = 3$ for magnetic dipole spin-down).  The
pulsar has roughly constant energy output until a time $\tau_0$,
beyond which the output declines fairly rapidly with time.

Based on studies of the Crab Nebula,
there appear to to be two electron populations - one corresponding
to relic radio-emitting electrons that may have been injected early
in the evolution of the supernova, and one associated with electrons
injected directly by the pulsar wind (Atoyan \& Aharonian 1996). These
are often parameterized as a broken power law:
\begin{equation}
Q(E_e,t) = Q_0(t)(E_e/E_b)^{-\alpha}
\end{equation}
with $\alpha = \alpha_1$ for $E_e < E_b$ and  $\alpha = \alpha_2$
for $E \ge E_b$. Here $E_b$ represents the energy of a spectral
break due to the two distinct populations. The integrated energy
in the electron spectrum is then
\begin{equation}
\int Q(E,t) E dE = (1 + \sigma) \dot{E}(t)
\end{equation}
where $\sigma$ is the ratio of the spin-down power injected in the form of
Poynting flux to that in the form of particles. 

The electrons injected into the PWN produce synchrotron radiation in the
nebular magnetic field, as well as inverse-Compton (IC) emission by
upscattering photons from the cosmic microwave background (CMB), ambient
starlight, and emission from nearby dust. The resulting emission spectrum
is found by integrating the electron spectrum over the emissivity function 
for synchrotron and inverse-Compton radiation (e.g. Lazendic et al. 2004)
using, respectively, the nebular magnetic field and spectral density of
the ambient photon field. For electrons of energy $E_{TeV}$ (in units
of TeV), the mean energy for synchrotron-emitted photons is
\begin{equation}
\epsilon_s \approx 2 \times 10^{-4} E_{TeV}^2 B_{-5} {\rm\ keV},
\end{equation}
where $B_{-5}$ is the magnetic field strength in units of $10^{-5}$~G,
while the average IC-scattered photon from the CMB has energy
\begin{equation}
\epsilon_{ic} \approx 3 \times 10^{-3} E_{TeV}^2 {\rm\ TeV}.
\end{equation}
For a typical PWN power law particle spectrum, the ratio of the synchrotron 
and IC fluxes
at these energies depends primarily on the magnetic field strength
(Aharonian \& Atoyan 1999):
\begin{equation}
\frac{f_{ic}(\epsilon_{ic})}{f_{s}(\epsilon_{s})} \approx 0.1 B_{-5}^{-2}
\end{equation}
where $f(E) = E^2 dF/dE$. The ratio of the peak $f(E)$ values for the 
spectra, shown schematically in Figure~2, is thus an indicator of the 
magnetic field strength.

\begin{figure}[b!]
 \resizebox{\columnwidth}{!}
  {\includegraphics{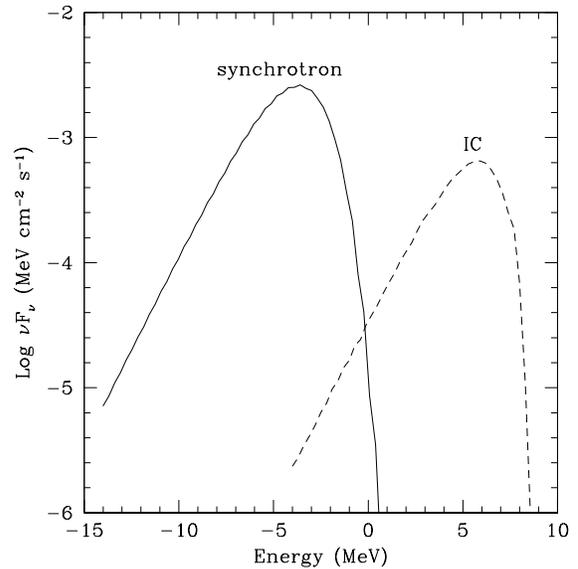}}
 \caption{
Illustration of the synchrotron and inverse-Compton emission from a
power law distribution of particles. Here we have used an exponential
cutoff at the high energy end of the particle distribution, a power
law index of 2.4, and magnetic field of $5 \mu$G. IC scattering is
from the CMB, and the particle spectrum normalization is arbitrary.
          }
\end{figure}

\begin{figure}[t]
 \resizebox{\columnwidth}{!}
  {\includegraphics{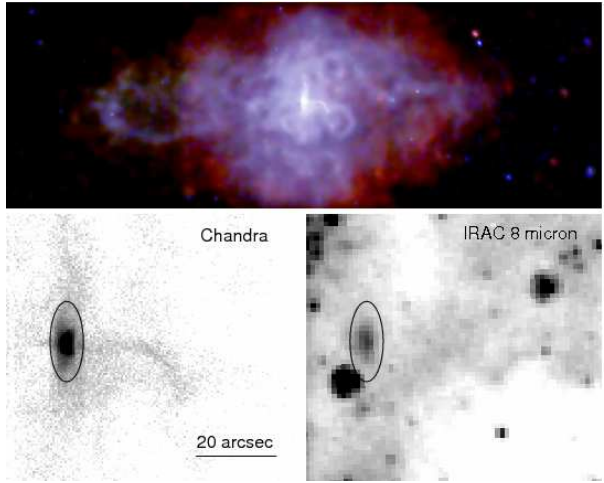}}
 \caption{
Upper: {\it Chandra} image of 3C~58 showing a complex of filamentary
structures embedded in a large-scale nebula. The pulsar is at the center,
accompanied by torus and jet-like structures.
Lower: {\it Chandra} image of the pulsar, torus, and jet (left), along
with {\it Spitzer} image revealing emission from the torus at
$8 \mu$m.
          }
\end{figure}

From the above, we see that for $B \sim 150 \mu$G, electrons that
produce 1~TeV photons from IC scattering of the CMB will produce
1~keV synchrotron photons.  For lower magnetic fields, the 
X-ray-emitting electrons are more energetic than the $\gamma$-ray-emitting
electrons. An important result is that for low magnetic fields and
large ages, synchrotron losses can result in a cooling break in the
electron spectrum that falls below the X-ray-emitting regime. The
result is a PWN that has very little X-ray emission, but is bright
in the VHE $\gamma$-ray band. This is discussed further below.

The evolution of the PWN spectrum is
determined by balance of injection and losses, 
\begin{equation}
\frac{dn_e(E_e,t)}{dt} = Q(E_e) - \frac{n_e(E_e,t)}{\tau_{rad}} 
- \frac{n_e(E_e,t)}{\tau_{esc}} 
\end{equation} 
where $\tau_{rad}$ and $\tau_{esc}$ are timescales for radiative and
escape losses, 
as well as the evolution of
the magnetic field. 
Venter \& de Jager (2006) parameterize the latter as
\begin{equation}
B(t) = \frac{B_0}{1 + (t/\tau_B)^\xi}
\end{equation}
where $\tau_B$ is a characteristic timescale connected to the
input history of the pulsar, and $\xi$ is a free parameter.

\begin{figure}
 \resizebox{\columnwidth}{!}
  {\includegraphics{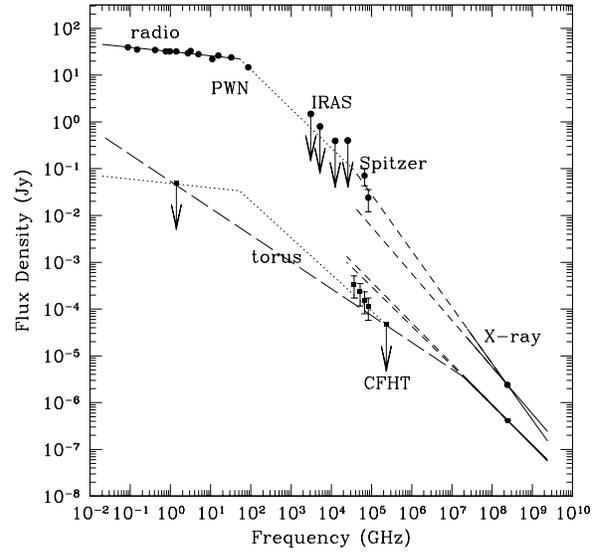}}
 \caption{
The flux of all of 3C~58 (upper) and its torus (lower), plotted
from the radio to the X-ray band. While the torus is not detected in the
radio band, the IR data require a flattening of the X-ray spectrum when
extrapolated back to the longer wavelength band.
          }
\end{figure}

Applying the above to the emission from a nebula of known age and
spin-down power provides constraints on the initial pulsar spin
period and magnetic field strength in the nebula, and has been used
successfully in modeling emission from a number of systems including
the Crab Nebula, the PWN associated with PSR B1509$-$58, and G0.9+0.1
(Venter \& de Jager 2006; Zhang et al. 2008).  It is important to
note that this describes a spatially-averaged spectrum for the
nebula, with no consideration given to variations of the magnetic
field, for example.  A more complete treatment of the nebula
structure, in which the particles emerge from a termination shock
in a the form of an axisymmetric particle and magnetic wind, which
is allowed to evolve through radiative and adiabatic losses in order
to determine the magnetic field distribution and electron spectrum,
is required to address spatial and spectral variations observed in
high resolution observations of PWNe. Recent work in this area has
revolutionized our understanding of the inner structure of PWNe
(e.g. Lyubarsky 2002, Komissarov \& Lyubarsky 2004), and extensions
to include the associated emissivity have now produced both detailed
emission maps and broadband spectral energy distributions based on
these more encompassing models (e.g. del Zanna et al., Volpi et al).

\subsection{A Point About Injection: 3C 58} While most attempts to
model the emission and evolution of PWNe start with the assumption
that a power law particle spectrum is injected into the nebula at
the wind termination shock, there is evidence that the spectrum may
be more complicated. Low-frequency spectral breaks are observed in
a number of PWNe, including 3C 58 and G21.5-0.9, for example. If
interpreted as cooling breaks, these would require uncomfortably
large magnetic fields (Green \& Scheuer 1992). Instead, it has been
suggested that these are the result of either an episodic change
in the pulsar input at an earlier epoch (e.g.  Woljter et al. 199x)
or an intrinsic feature in the injection spectrum (Frail 1998).

{\it Chandra} Observations of 3C 58 (Slane et al. 2004) reveal an
extended nebula filled with a complex of loop-like structures and
surrounded by a shell of swept-up ejecta (Figure 3). At the center
is an energetic young pulsar accompanied by a jet and surrounded
by a toroid that is also observed with {\it Spitzer} observations
(Slane et al. 2008). These observations, combined with data from
other wavebands, show that the spectrum emerging from the wind
termination shock in this system requires one or more spectral
breaks (Figure~3).  Such breaks will imprint themselves on the
broadband spectra of PWNe, causing the emission to deviate from
that expected from a pure power law input, and potentially producing
the low energy breaks observed for the large scale nebula, such as
that seen in 3C~58 (Figure~3).  Consideration must be given to such
complexities in the injection spectrum to properly understand the
evolution of the broadband spectrum.

\section{Evolution of PWNe Within SNRs}

Since a pulsar is formed in a supernova explosion, the star and its
PWN are initially surrounded by the expanding ejecta and swept-up
circumstellar material that comprise the associated SNR. At early
times the SNR blast wave expands rapidly ($v_s > (5-10)\times10^3
{\rm\ km\ s}^{-1}$), while the pulsar itself moves at a slower (but
often significant) speed (with typical magnitude $\sim 400 {\rm\
km\ s}^{-1}$) as the result of a ``kick'' due to asymmetries in the
explosion.  In the early phases of evolution, the pulsar is thus
located very near the center of the SNR.

The energetic wind is injected into the SNR interior, forming a
high-pressure bubble that expands supersonically into the surrounding ejecta,
forming a shock. The radius of the PWN
evolves as 
\begin{equation}
R_{PWN} \approx 1.5  \dot{E}_0^{1/5} E_{SN}^{3/10} M_{ej}^{-1/2} t^{6/5}
\end{equation}
where $E_{SN}$ is the energy released in the explosion and $M_{ej}$
is the mass of the ejecta (Chevalier 1977). Thus, at least at early 
times when the
pulsar input is high, the PWN expansion velocity increases with
time. The sound speed in the relativistic fluid within the PWN is
sufficiently high ($c_s = c/\sqrt{3}$) that any pressure variations
experienced during the expansion are quickly balanced within the
bubble; at early stages, we thus expect the pulsar to be located
at the center of the PWN. The pressure balance within the PWN results
in a termination shock where the energetic pulsar wind meets the
more slowly-expanding PWN.

The outer blast wave of the SNR, meanwhile, drives a shock into the
surrounding interstellar/circumstellar medium (ISM/CSM), forming a
shell of hot gas and compressed magnetic field. As the shell sweeps
up additional mass, and decelerates, a reverse shock (RS) propagates
back into the expanding ejecta. As illustrated in Figure 1, the
entire system is thus described by four shocks: a forward shock that
sweeps up the CSM/ISM; a reverse that heats the supernova ejecta
(which is initially cold due to severe adiabatic losses from the
rapid expansion); a shock at the PWN/ejecta boundary, where the
expanding nebula compresses and heats surrounding ejecta; and the
pulsar wind termination shock. At early times, the SNR reverse shock
expands outward, but more slowly than the forward shock, but eventually
it moves inward.
In the absence
of a central pulsar or PWN, the reverse shock reaches the center
of the SNR in a time
\begin{equation}
t_{c} \approx 7 \left( \frac{M_{ej}}{10~M_\odot}\right)^{5/6}
\left( \frac{E_{SN}}{10^{51}~{\rm ergs}} \right)^{-1/2}
\left( \frac{n_0}{{\rm cm}^{-3}} \right)^{-1/3}~{\rm kyr},
\end{equation}
where $n_0$ is the number density of ambient gas ((Reynolds \& Chevalier
1984).

\section{RS/PWN Interactions}

In the presence of a young pulsar, the reverse shock collides with the
expanding PWN at a time
\begin{equation}
t_{col} \approx 1045 E_{51}^{-1/2} \left(\frac{M_{ej}}{M_\odot}
\right)^{5/6} n_0^{-1/3} {\rm\ yr}
\end{equation}
where $E_{51}$ is the explosion energy in units of $10^{51}$~erg 
(van der Swaluw et al. 2004). This 
compresses the PWN until the pressure in the nebula
is sufficiently high to rebound, and again expand into the ejecta.
The system goes through several reverberations over the course of
several tens of thousands of years.

The crushing of the PWN results in an increase in the magnetic field.
This causes a rapid burn-off of the most energetic particles in
the nebula. The PWN/RS interface is Rayleigh-Taylor unstable, and
is subject to the formation of filamentary structure where the
dense ejecta material is mixed into the relativistic fluid (Blondin
et al. 2001). The nebula subsequently re-forms as the pulsar 
injects fresh particles into its surroundings, but a significant
relic nebula of mixed ejecta and relativistic gas will persist.

In cases where the pulsar and its PWN have moved considerably
from the center of the PWN, or in which the SNR has evolved in
a medium of nonuniform density, the reverse shock will interact
with the PWN asymmetrically, encountering one portion of the nebula
well before another. This results in a complex interaction that leaves a 
highly distorted relic nebula that may be highly displaced from
the pulsar position (Blondin et al. 2001).

\section{PWNe as Extended VHE $\gamma$-ray Sources}

Recent observations with the latest generation of VHE $\gamma$-ray
telescopes have uncovered a large number of sources that are, or
appear to be, associated with PWNe. As described above, the broadband
spectra provide strong constraints on the structure and magnetic
fields in these systems.  Coupled with information about their
angular sizes, distances, and associations with any identified
compact objects or supernova remnants, these observations are
providing new insights into PWN evolution. Here I summarize results
from three such studies that illustrate some of the key results
that have been obtained.

\begin{figure}[p]
  \includegraphics[height=0.9\textheight]{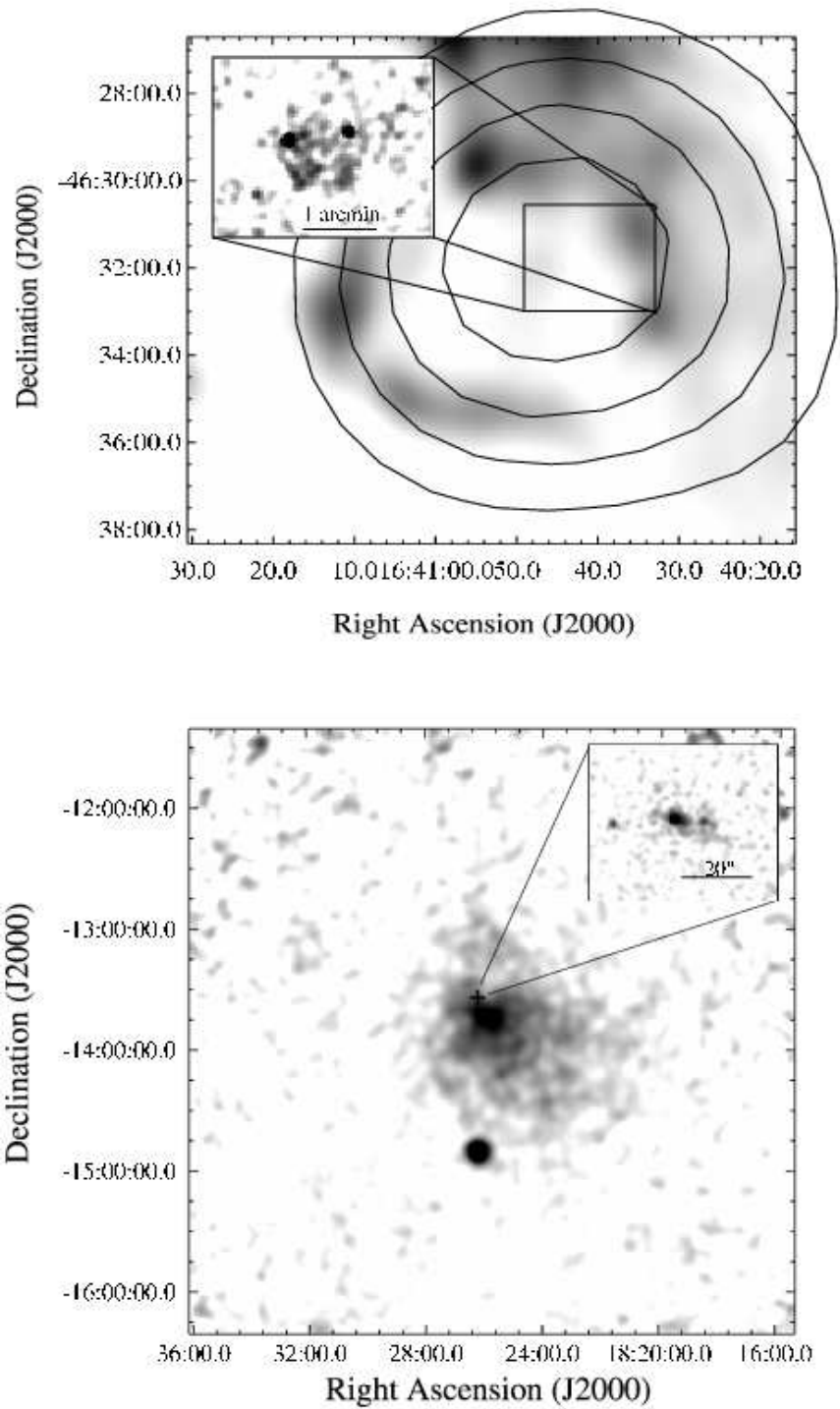}
  \caption{
Upper: Radio image of G338.3$-$0.0 with contours from HESS J1640$-$465. The
inset shows the {\it Chandra} image of the central regions, revealing
the putative pulsar (the brightest source in the field) surrounded by
a faint extended X-ray nebula.
Lower: H.E.S.S. image of J1825$-$137. The cross marks the position of
PSR B1823$-$13 and the inset shows the {\it Chandra} image of
the pulsar, revealing a faint extended nebula.
}
\end{figure}

\subsection{HESS J1640$-$465}

HESS J1640-465 (Aharonian et al. 2006a) was identified as an extended
source of VHE $\gamma$-ray emission in a survey of the inner Galaxy
carried out by {\it H.E.S.S.} in 2004-2006. The source has a radius
of $\sim 3$~arcmin, and the spectrum is well-described by a power
law with an index of $\sim 2.4.$ The source is spatially coincident
with G338.3$-$0.0, a shell-type SNR with a diameter of $\sim 8$~arcmin
(Figure 5, top). The peak of the $\gamma$-ray emission falls interior
to the radio shell, suggesting that the emission may be from a
pulsar-powered nebula associated with the SNR, although no known
pulsar resides in this vicinity.

X-ray observations with {\it XMM-Newton} (Funk et al. 2007) reveal
extended X-ray emission in the central regions of HESS J1640-465,
suggestive of a PWN core. Subsequent {\it Chandra} observations
(Lemiere et al. 2008) reveal a compact object surrounded by a
faint diffuse nebula (Figure 5, top inset) whose nonthermal X-ray spectrum 
shows clear evidence for steepening with radius, as expected if
strong synchrotron losses have prevented energetic electrons
from diffusing to large distances from the putative pulsar. 

HI absorption measurements (Lemiere et al. 2008) establish a distance
of 8 - 13~kpc for G338.3$-$0.0, suggesting an age for the system
of 10 - 30~kyr.  Based on the X-ray luminosity of $\sim 4 \times
10^{33} d_{10}^2 {\rm\ erg\ s}^{-1}$ for the X-ray nebula (where
$d_{10}$ is the distance in units of 10~kpc), a typical efficiency
of $\sim 0.1\%$ for conversion of spin-down power to X-ray emission
suggests $\dot{E} \sim 4 \times 10^{33} {\rm\ erg\ s}^{-1}$ for the
putative pulsar.  Using the approach described in Section 2, a
leptonic model can adequately describe the X-ray and $\gamma$-ray
spectra, and predicts a system age of $\sim 15$~kyr
with a mean magnetic field strength of $\sim 6 \mu$G.

The observed position and morphology of the X-ray nebula
associated with HESS J1640$-$465 is somewhat perplexing. It is not
at the center of the SNR shell, and the extended emission is
not symmetric about the position of the putative pulsar. If
born at the geometric center of G338.3$-$0.0, its current position
indicates a projected space velocity of $\sim 500 {\rm\ km\ s}^{-1}$,
which is not unreasonable for a pulsar. However, the asymmetric
shape of the diffuse nebula makes such an explanation problematic.
A more likely scenario may be that the PWN has already undergone
disruption from the RS interaction. Based on the distribution and
kinematics of the molecular and neutral gas in this vicinity,
it appears likely that the SNR has evolved in a sufficiently 
complex environment to expect an asymmetric RS/PWN interaction
(Lemiere et al. 2008).

\subsection{HESS J1825$-$137}

Also discovered as part of the {\it H.E.S.S.} Galactic plane survey, HESS
J1825$-$137 is a large-diameter ($\sim 1^\circ$) $\gamma$-ray source
with a highly asymmetric morphology (Figure 5, bottom). It is apparently
associated with the PSR B1823$-$13, located at a distance of $\sim
4$~kpc. The pulsar has a spin-down power $\dot{E} = 2.8 \times
10^{36}{\rm\ erg\ s}^{-1}$ and is surrounded by a faint X-ray nebula
(see Figure 5, bottom inset) whose extended morphology, similar in
orientation to that of HESS J1825$-$137 but on a much smaller scale,
suggests that the PWN has been disrupted by the SNR reverse shock
(Gaensler et al. 2003). This is consistent with the significant
offset between the pulsar and the geometric center of HESS J1825$-$137.

Remarkably, the radius of HESS J1825$-$137 is observed to decrease
with increasing energy in the $\gamma$-ray band, providing the
first evidence in this band for the effects of synchrotron burn-off on the 
underlying  electron spectrum. Variations in the emission 
spectrum confirm a spectral steepening with radius, as expected
from such losses. These observations imply a magnetic field strength
of only $\sim 2-4 \mu$G, and an age of $\sim 20-40$~kyr (de Jager
\& Djannati-Ata\"i). 

\begin{figure}[t!]
  \includegraphics[height=0.6\textheight]{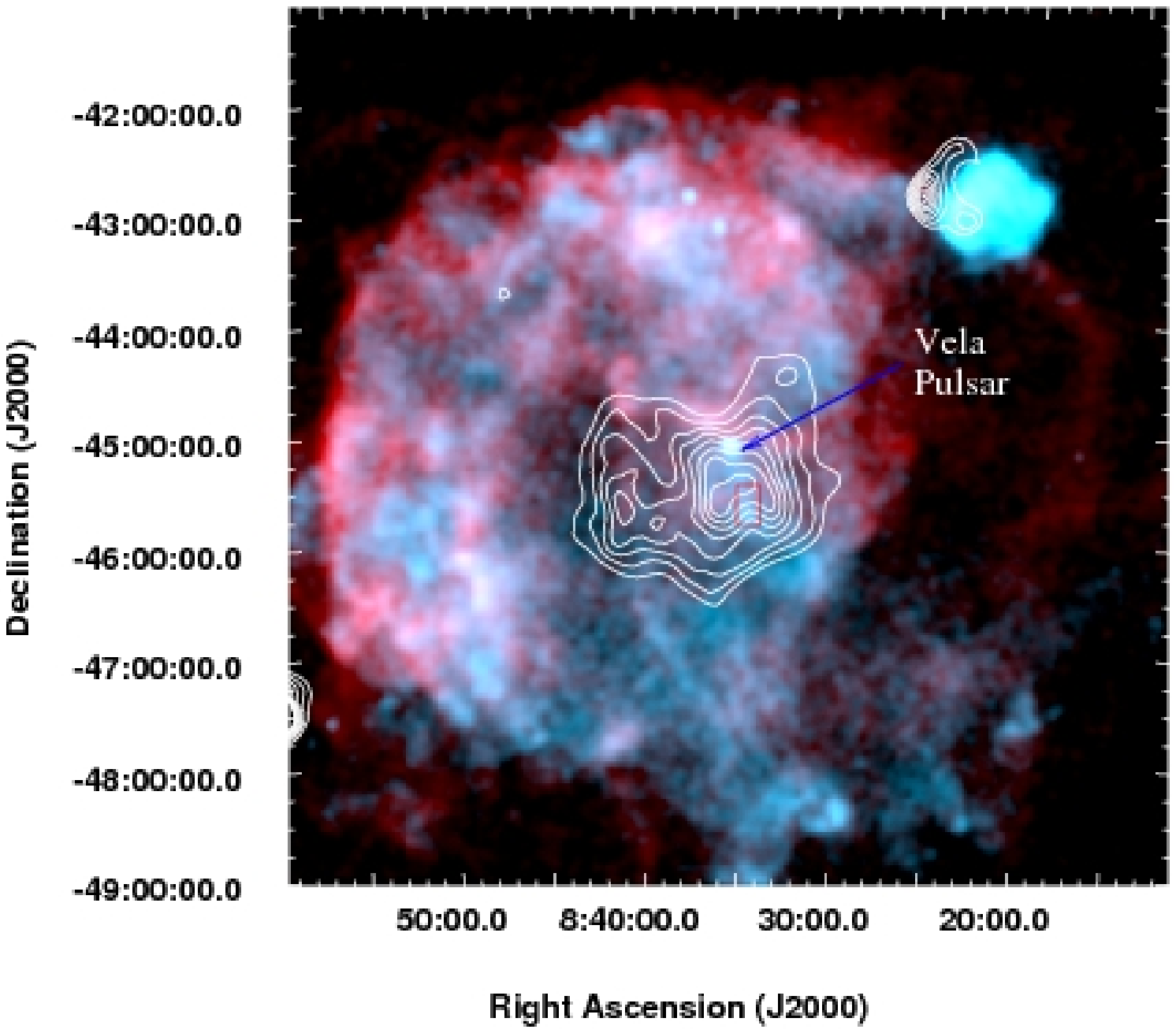}
  \caption{{\it ROSAT} image of the Vela SNR. Radio emission from
the associated PWN Vela X is shown with contours. The pulsar is
indicated with an arrow, and is offset from the center of Vela X,
presumably due to the reverse shock interaction. Hard X-ray emission,
shown in blue, extends southward from the pulsar, into Vela X, and
this structure is accompanied by VHE $\gamma$-ray emission. The red
box indicates the region from which {\it XMM-Newton} spectra were
extracted (see text and Figure 7).
}
\end{figure}

The extremely large size of HESS J1825$-$137 makes this the largest
PWN known. With a radius of nearly 35~pc, it appears clear that the
system has evolved well past the reverse-shock crossing phase, and
that the PWN has rebounded and continued to expand within its unseen
host SNR. The spectral steepening seen in the {\it H.E.S.S.} data
suggest that the radiative loss break has progressed to energies
well below X-ray-emitting values, thus explaining the lack of
X-ray emission from sites other than very near the pulsar, where fresh
particle injection is taking place. The properties of HESS J1825$-$137
suggest that there may be a significant population of $\gamma$-ray-bright
sources with very faint or undetectable counterparts -- perhaps
providing an explanation for the emerging number of ``dark'' TeV sources
(de Jager \& Djannati-Ata\"i 2008).

\subsection{Vela X}
The Vela supernova remnant (Figure 6) is a prototype of the composite class
in which the shell of hot gas swept up by the forward shock of the
explosion is accompanied by a nonthermal nebula driven by the pulsar
formed in the collapse of the progenitor star. The SNR shell is
quite cool, indicating an object of moderate age, and fragments of
metal-rich ejecta that have overtaken the decelerated blast wave
are observed well outside the shell. The pulsar itself is accompanied
by a compact jet-torus structure that provides the direction of the
pulsar spin axis, and the extended wind nebula -- identified with
the large radio region known as Vela~X --  is located predominantly
to the south of the pulsar, apparently having been disrupted by the
reverse shock in the remnant.

\begin{figure}[t]
  \includegraphics[width=.51\textwidth]{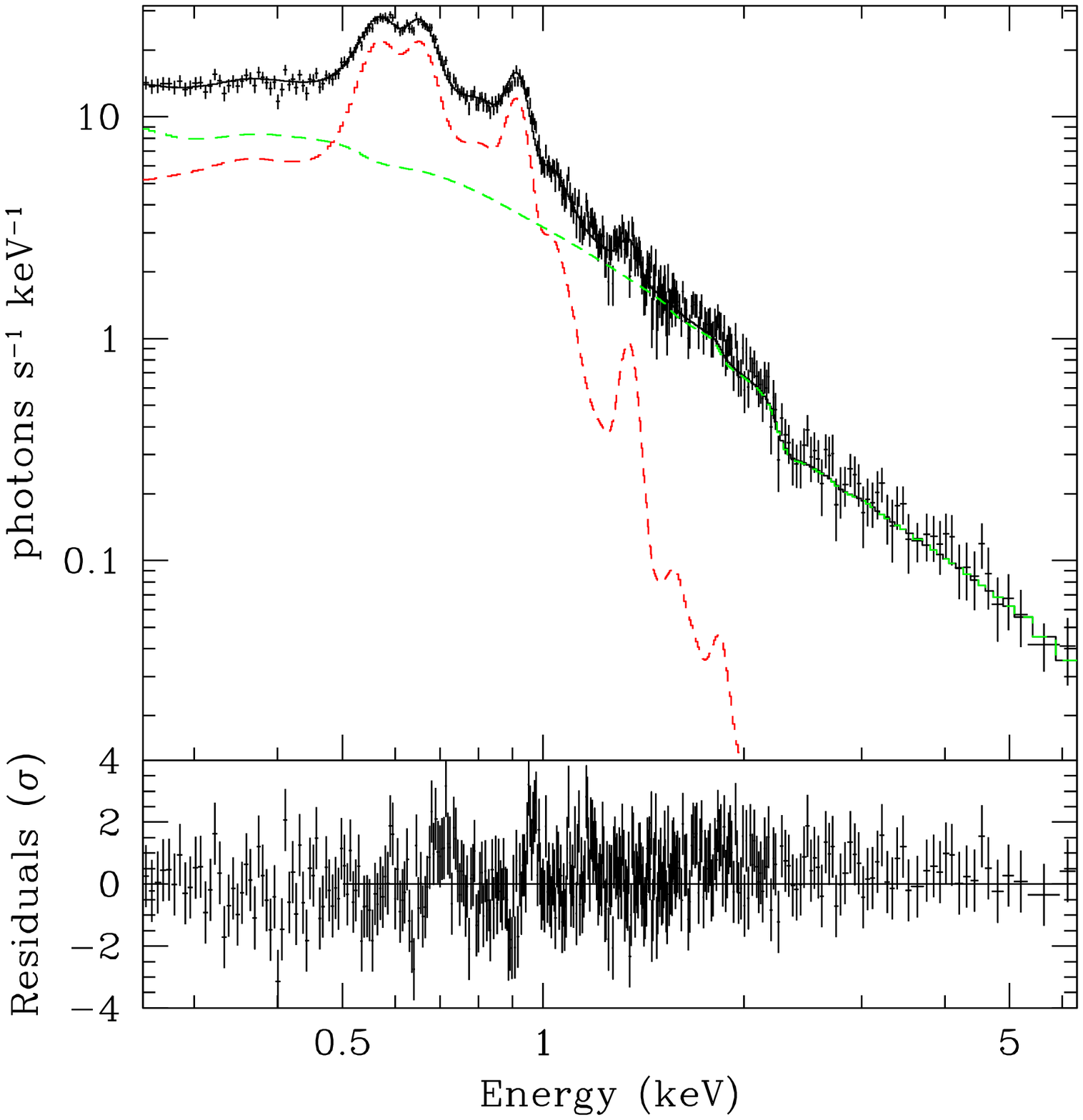}
  \includegraphics[width=.51\textwidth]{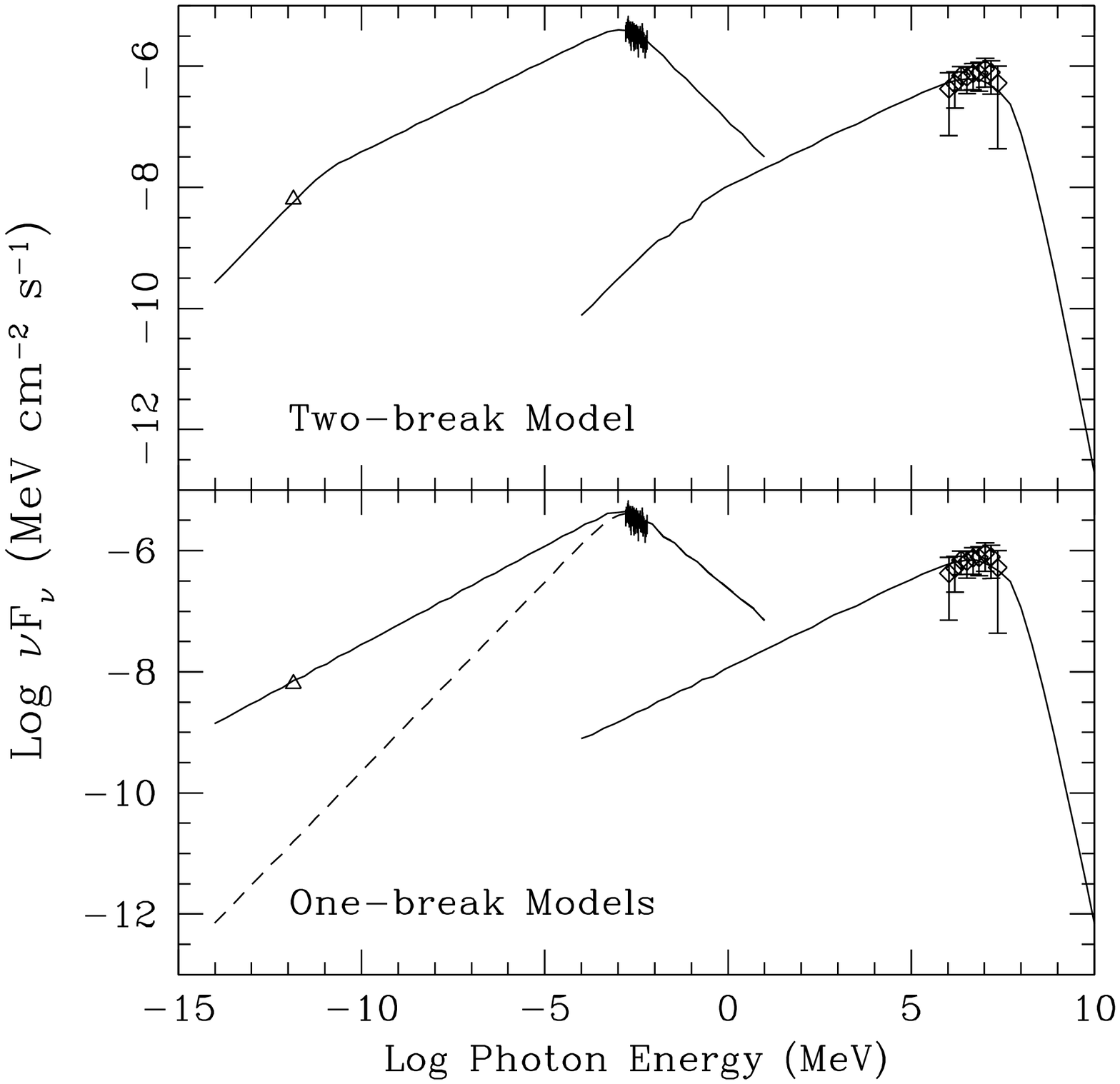}
  \caption{
Left: {\it XMM-Newton} spectrum from the ``cocoon'' region in Vela X.
The best-fit model, shown in black, is composed of two components --
a thermal plasma with enhanced, ejecta-like abundances (shown in red)
and a power law (shown in green).
Right: Broadband spectral model consisting of synchrotron emission
in the radio and X-ray bands accompanied by inverse-Compton emission
in the VHE $\gamma$-ray band. The upper panel shows a model with
two spectral breaks in the electron spectrum. Models with a single break
(lower panel) either underpredict the radio emission, or produce a
radio flux whose spectral index does not agree with observations.
(From LaMassa et al. 2008.)
}
\end{figure}

ROSAT observations of the Vela~X region (Markwardt \& \"{O}gelman
1995) reveal a large emission structure extending to the south of
the pulsar (Figure 8). The region is characterized by a hard 
spectrum and appears to lie along a bright elongated radio structure
(Frail et al. 1997). Observations with {\it H.E.S.S.}
reveal bright VHE $\gamma$-ray emission from a region extending directly
along this feature (Aharonian et al. 2006b). 

Spectral studies carried out with XMM-Newton (LaMassa et al. 2008)
reveal two distinct emission components in the X-ray band -- a power
law with a spectral index of $\sim 2.2$ and a thermal plasma with
enhanced abundances of O, Ne, and Mg, presumably associated with
ejecta that has been mixed into the PWN upon its interaction with
the reverse shock.  The derived density of the thermal-emitting,
ejecta-rich gas is $n \sim 0.06 {\rm\ cm}^{-3}$, too low to accommodate
hadronic scenarios for the $\gamma$-ray emission.

A leptonic model consisting of a broken power law for the electrons
in this region of the nebula is able to successfully reproduce the
broadband emission as a combination of synchrotron radiation in the
radio and X-ray bands, and inverse-Compton scattering of the cosmic
microwave background from the same electron population to produce
the $\gamma$-ray emission (Figure 8). The best-fit model requires
a magnetic field strength of $\sim 5 \mu$G and
 an electron spectrum with two spectral breaks.
The lower break is presumably intrinsic to the injected electron
spectrum, or corresponds to a distinct population of radio-emitting
particles injected early in the formation of the PWN, while the
upper break is presumably associated with synchrotron cooling of
the electrons.  Models with a single break are also possible, but
violate either the observed radio flux or its spectral index 
(LaMassa et al. 2008)

Our understanding of the low-energy electron spectrum in Vela X
is severely limited by the lack of observations at frequencies 
beyond about 30~GHz. It is unclear whether the spectrum has
a break in this region, as suggested above, or continues to
higher energies as a distinct population. 
However, electrons that would produce emission beyond the current
radio frequency limits will also upscatter ambient photons into
the energy sensitivity range of the {\it Fermi} LAT (de Jager et al.
2008). Indeed, modeling shows that the three distinct photon
populations (the CMB, ambient starlight, and IR emission from ambient
dust) will produce discrete and resolvable features in the LAT.
Future results from the ongoing LAT sky survey thus promise to
address key questions about the particle distribution in Vela~X.
This, combined with more extensive studies of the thermal and nonthermal
X-ray emission will place strong constraints on the full evolution
of Vela~X, from the formation stage through the process of reverse-shock
crushing of the PWN.

\section{Summary}
The broadband spectra of PWNe provide information about both the
structure and evolution of these objects. New measurements in
the VHE $\gamma$-ray band, combined with detailed measurements
provided by radio and X-ray studies, have begun to reveal information
about the earliest stages of particle injection in PWNe as well as
a view to the late-phase evolution characteristics. These sources
currently represent a significant fraction of the known VHE 
$\gamma$-ray sources, and future observations, combined with those
from {\it Fermi} are sure to extend this new frontier. Complemented
by multiwavelength observations, these results promise to extend
the current flourish of activity in observational and theoretical
work on pulsar winds.


\begin{theacknowledgments}
The author acknowledges significant contributions to this work by
B. Gaensler, S. Reynolds, D. Helfand, J. Gelfand, O. C. de Jager, 
S. LaMassa, and A. Lemiere.
Partial support was provided by NASA Contract
NAS8-03060.
\end{theacknowledgments}


\begin{thebibliography}{9}

\bibitem{aa99}
F. Aharonian and A.~M. Atoyan \emph{A\&A} \textbf{351}, 330 (1999).

\bibitem{ahess06}
F. Aharonian, et al.,  \emph{A\&A} \textbf{460}, 489 (2006a).

\bibitem{ahess06b}
F. Aharonian, et al.,  \emph{A\&A} \textbf{448}, L43 (2006b).

\bibitem{bcf01}
J.~M. Blondin, R.~A. Chevalier, and D.~M. Frierson, \emph{ApJ}
\textbf{563}, 806 (2001).

\bibitem{che77}
R.~A. Chevalier, \emph{In Supernovae, ed. DN Schramm, pp. 53-61.
Dordrecht:Reidel} (1977).

\bibitem{dd08}
O.~C. de Jager and A. Djannati-Ata\"i, \emph{arXiv:0803.0116v1} (2008).

\bibitem{dsl08}
O.~C. de Jager, P. Slane, and S.~M. LaMassa, \emph{ApJ} in press (2008).

\bibitem{dvab06}
L. Del Zanna, D. Volpi, E. Amato, and N. Bucciantini, \emph{A\&A} 
\textbf{453}, 621 (2006).

\bibitem{frail98}
D.~A. Frail,  \emph{In ``The Many Faces of Neutron Stars.'' Edited by
R. Buccheri, J. van Paradijs, and M. A. Alpar. Dordrecht ; Boston : Kluwer
Academic Publishers, 1998., p.179} (1998).

\bibitem{frail97}
D.~A. Frail et al.,  \emph{ApJ} \textbf{475}, 224 (1997).

\bibitem{funk07}
S. Funk et al.,  \emph{ApJ} \textbf{267}, 517 (2007).

\bibitem{gaensler et al 2003}
B.~M. Gaensler et al.,  \emph{ApJ} \textbf{588}, 441 (2003).

\bibitem{GS06}
B.~M. Gaensler and P.~O. Slane, \emph{ARA\&A} \textbf{44}, 17 (2006).

\bibitem{GS92}
D.~A. Green and P.~A.~G. Scheuer, \emph{MNRAS} \textbf{258}, 833 (1992).

\bibitem{h08}
J.~J. Hester, \emph{ARA\&A} \textbf{46}, 127 (2008).

\bibitem{kc84a}
C.~F. Kennel  and F.~V. Coroniti, \emph{ApJ} \textbf{283}, 694 (1984).

\bibitem{kl04}
S. Komissarov and Y.~E. Lyubarsky, \emph{Ap\&SS} \textbf{293}, 107 (2004).

\bibitem{lsd08}
S.~M. LaMassa, P.~O. Slane, and O.~C. de Jager, \emph{ApJ} in press (2008).

\bibitem{lsetal04}
J.~S. Lazendic et al., \emph{ApJ} \textbf{602}, 271 (2004).

\bibitem{lemiereetal08}
A. Lemiere et al., \emph{ApJ} submitted (2004).

\bibitem{yel02}
Y.~E. Lyubarsky, \emph{MNRAS} \textbf{329}, L34 (2002).

\bibitem{mo95}
C. Markwardt and H. \"{O}gelman, \emph{Nature} \textbf{375}, 40 (1995).

\bibitem{PS73}
F. Pacini and M. Salvati, \emph{ApJ} \textbf{186}, 249 (1973).

\bibitem{RC84}
S.~P. Reynolds, and R.~A. Chevalier, \emph{ApJ} \textbf{278}, 630 (1984).

\bibitem{shvm04}
P. Slane et al., \emph{ApJ} \textbf{616}, 403 (2004).

\bibitem{slaneetal08}
P. Slane et al., \emph{ApJ} \textbf{676}, L33 (2008).

\bibitem{vd06}
C. Venter and O.~C. de Jager, \emph{astro.ph.12652} (2006).

\bibitem{vdk04}
E. van der Swalum, T.~P. Downes, and R. Keegan,, \emph{A\&A} \textbf{420},
937 (2004).

\bibitem{vdab08}
D. Volpi, L. Del Zanna, E. Amato, and N. Bucciantini, \emph{A\&A}
\textbf{485}, 337(2008).

\bibitem{wspb97}
L. Woltjer, M. Salvati, F. Pacini, and R. Bandiera, \emph{A\&A} \textbf{325},
295 (1997).

\bibitem{zc08}
L. Zhang, S.~B. Chen, and J. Fang, \emph{ApJ} \textbf{676}, 1210 (2008).

\end{thebibliography}
\end{document}